# Properties of ZnO/ZnAl$_2$O$_4$ composite PEO coatings on zinc

M. Serdechnova[1], C. Blawert[1], S. Karpushenkov[2], L. Karpushenkava[2], T. Shulha[1], P. Karlova[1], R. Vasilić[3], S. Stojadinović[3], S. Stojanović[4], Lj. Damjanović-Vasilić[4], V. Heitmann[1], S.M. Rabchynski[2], M.L. Zheludkevich[1,5]

[1] Institute of Materials Research, Helmholtz-Zentrum Geesthacht, Max-Planck-Straβe 1, 21502 Geesthacht, Germany

[2] Belarusian State University, Faculty of Chemistry, Nezavisimosti Avenue 4, 220030 Minsk, Belarus

[3] University of Belgrade, Faculty of Physics, Studentski trg 12-16, 11000 Belgrade, Serbia

[4] University of Belgrade, Faculty of Physical Chemistry, Studentski trg 12-16, 11000 Belgrade, Serbia

[5] Institute for Materials Science, Faculty of Engineering, University of Kiel, Kaiserstraße 2, 24143 Kiel, Germany

**Abstract**

Recently the successful formation of PEO coatings on zinc in a phosphate aluminate electrolyte was shown. The produced composite coatings contain various mixtures of ZnO and ZnAl$_2$O$_4$. In frame of the current study, the properties of the formed coatings including adhesion/cohesion, wear, corrosion and photocatalytic activity were analysed to identify possible applications. However, the coatings show internal porosity and a sponge-like structure. Thus the cohesion within the coating is quite low. Pull-off tests have demonstrated clear rupture within the PEO layer at strength values as low as 1 MPa. The photocatalytic activity is limited, in spite of the formation of a higher amount of ZnO at shorter treatment times. Interestingly, the composite coatings of ZnO and higher amounts of ZnAl$_2$O$_4$ spinel showed a higher activity, but not sufficient for fast and effective catalytic cleaning applications.

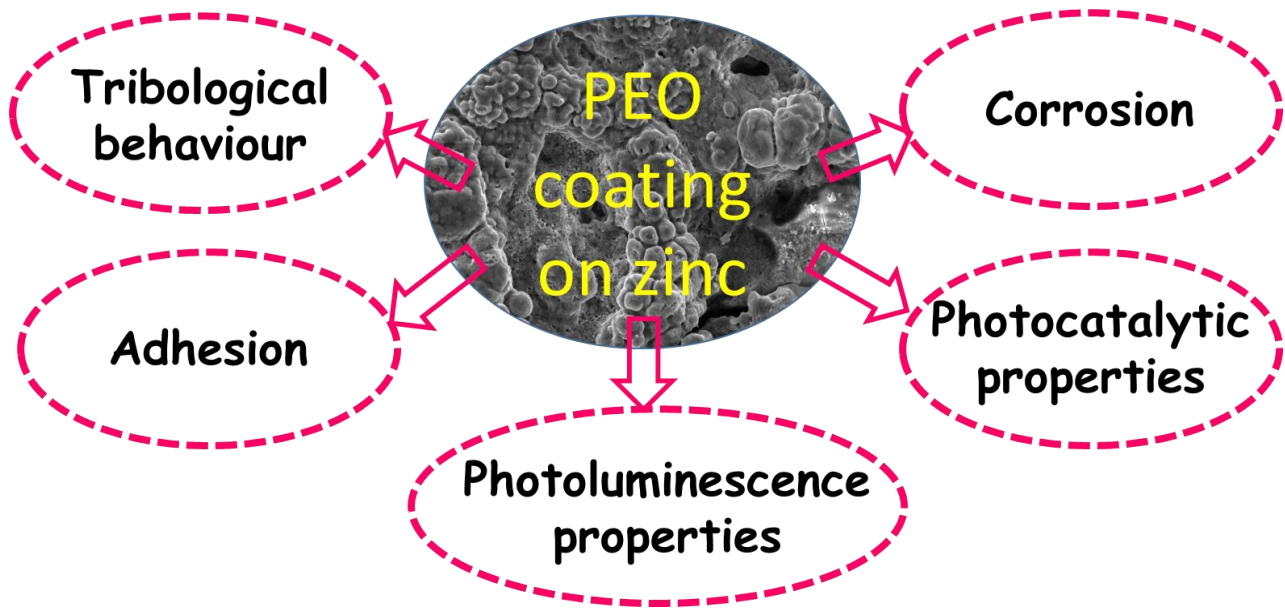

**Highlights**

PEO coatings consisting mainly of $ZnAl_2O_4$ are synthesized on a Zn alloy

The coatings reveal a sponge like microstructure with micro porosity.

In spite of the porosity, improvements of wear and corrosion resistance are achieved.

# Introduction

Plasma electrolytic oxidation (PEO) is an advanced anodizing process, which leads to the formation of ceramic-like coatings on the surface of light and valve metals [1-3]. This happens as a result of short-living microdischarges (milliseconds range), which take place under high voltages in low-concentrated, environmentally friendly electrolytes [4-7]. The oxide layers developed by PEO method are usually hard and well adherent to the substrates and can provide effective corrosion and wear resistance [8-12]. Thus, PEO treatments are well known for corrosion and wear protection of magnesium [5, 7, 15-17], aluminium [18-23] and titanium [24-29] based alloys.

PEO layers typically reveal a layered structure, where a harder outer porous layer is responsible for the wear resistance, while an inner denser layer confers the corrosion resistance [5, 30-35]. The porous morphology of the PEO layers on the one hand compromises the barrier properties of the formed coatings, but on the other hand offers an opportunity for additional sealing [36-38] and for effective, further surface functionalization [39-41]. For example, the PEO pores can be helpful for the adhesion of polymer top-coats due to the improved mechanical interlocking [42-44], they can be loaded with suitable corrosion inhibitors [39, 45, 46] or sealed with other functional species [47, 48].

However, too many pores can result in the formation of a pore band which is often located between the outer layer and the inner barrier layer and can have negative effects on the cohesion of the coating if a crack can easily propagate along the pore band [49, 50]. This can lead to local cohesive failure of the coating and subsequent damage the whole coating system. A pore band often is formed in pulsed DC PEO processing of Mg alloys [5], but just recently a similar, but even more severe form was observed for PEO treated Zn alloy [51].

Zinc also belongs to the materials, on which formation of PEO coatings is possible [9, 51-53]. For example, a flash PEO treatment with post-sealing was performed for corrosion protection [9], formation of ZnO by PEO processing of Zn was studied because of its interesting combination of optical and electrical properties [53] as well as possible biomedical applications [52]. However, the total amount of studies regarding PEO of Zn alloys is very limited [52] and consequently also the knowledge about coating and phase formation in different electrolytes, coating properties and possible applications. Recently, a first study performed in an aluminate/phosphate based electrolyte was published by Blawert et al. [51]. Focus was on the coating growth kinetics as a function of different current densities, surface morphology, elemental distribution and phase composition of the coatings. The mechanism of PEO coating formation was suggested, but no attention was paid on the coating properties. Stresses in the coatings were identified, causing flaking off of the coatings. To prevent this, a special kind of "repair" processing treatment was applied. However, general and functional properties as well as the corrosion protection capability of the obtained coatings were not targeted and reported in that mechanistic study. One of the expected functionality is attributed to the potential photocatalytic properties on the basis of ZnO present in the layers.

Thus the aim of the current study is to analyse the properties of the formed coatings in detail regarding coating cohesion, wear, corrosion and photocatalytic activity and to identify possible fields of applications for the PEO coatings on zinc.

**Experimental**

*Materials*

For electrolyte preparation and PEO synthesis deionized water was used as a solvent for the following chemicals: potassium hydroxide (KOH, >99 %, Sigma-Aldrich Chemie GmbH, Germany), trisodium phosphate ($Na_3PO_4$, techn., Alfa Aesar, Germany), sodium aluminate ($NaAlO_2$, techn., Riedel-de Haën, Germany). Zinc alloy sheet with 1 mm thickness (Z1, according to EN988,) and a nominal composition of [wt.%]: 0.08 - 1.00% Cu, 0.06 - 0.20% Ti, ≤ 0.015% Al and Zn balance was purchased from VMZinc. From the sheet, specimens with various dimensions were tooled (20 x 25 x 1 for all characterisations, except 50 x 25 x 1 and 50 x 50 x 1 for adhesion testing). The specimens were ground with 2500 grit silicon carbide abrasive paper and cleaned with ethanol before PEO treatment. The photocatalytic activities were evaluated by degrading test dye molecule methyl orange (MO) (C.I. 13025 indicator ACS, Merck, Germany).

*Methods*

PEO processing was performed using a pulsed DC power supply under a constant voltage limit of 450 V and with a pulse ratio equal to $t_{on}$: $t_{off}$ = 1ms: 9ms. The aqueous electrolyte contained 1g $L^{-1}$ KOH, 8 g $L^{-1}$ $Na_3PO_4$ and 12g $L^{-1}$ $NaAlO_2$. The treatment was performed under continuous stirring at 20 ± 2 °C. Latter was maintained by a circulating external water cooling system. Four different current densities (75, 100, 125 and 150 mA $cm^{-2}$) and three different treatment times (3, 6 and 12 min) were investigated. The prepared specimens were rinsed with deionized water and dried under air.

*Surface morphology* as well as cross sectional views of PEO treated samples were examined using *Tescan Vega3 SB* scanning electron microscope (SEM, Brno, Czech Republic) equipped with an *eumeX* energy dispersive X-ray (EDS, Heidenrod, Germany) spectrometer. SE mode was selected for analysis of morphology of original PEO coatings, while BSE UniVac mode was used in order to analyse the sliding wear tracks.

The integral *phase composition* of the coatings was analysed using X-ray diffraction (XRD, Bruker D8 Advance, Ni filtered Cu $K_α$ radiation source). A step size of $0,02^O$ and a scan range from 15 $^O$ till $50^O$ was settled, with a glancing angle of $3^O$ and a scan speed of 2 s per step.

The adhesion/cohesion of the various PEO coatings was assessed by pull-off tests. Aluminium dollies (20 mm diameter) were glued on the PEO coated zinc surface using Araldite 2011. The glue dried at ambient conditions for 22 hours and the coating was removed around the dolly for obtaining a defined testing diameter prior to the measurements. Pull-off tests were performed using POSITEST AT-M manual adhesion tester (DeFelsko, NY, USA).

The *dry sliding wear behaviour* of the PEO coatings was assessed with an oscillating ball-on-disc tribometer (Tribotec AB, Clichy-France, France), with an AISI 52100 steel ball of 6 mm diameter as the static friction counterpart (IHSD-Klarmann, Bamberg, Germany). The wear tests were performed at ambient conditions (25 ± 2 °C and 36–44% relative humidity) with different loads ranging from 1 to 10 N and an oscillating amplitude of 10 mm with a sliding velocity of 5 mm s$^{-1}$. The test was terminated after a total sliding distance of 12 m. Laser scanning confocal microscope (LSM 800, ZEISS, Germany) was used for analysis of the wear tracks after the test. ConfoMap$^{ST}$ software was used for subsequent data treatment and analysis.

*Electrochemical experiments* were performed using a Gamry Interface 1000 potentiostat (Gamry, Warminster, USA) and a conventional three-electrode cell, employing a platinum wire as counter electrode and a silver/silver chloride reference electrode (Ag/AgCl) showing a potential of 0.210 V with respect to the standard hydrogen electrode (SHE). As working electrode the coated material (exposed area 0.5 cm$^2$) was used and the test solution was naturally aerated 0.5 wt.% NaCl electrolyte at room temperature (22 °C). For the measurements the electrochemical cell was placed in a Faraday cage. All impedance spectra were recorded at open-circuit potential, applying a sinusoidal perturbation of 10 mV RMS amplitude and a frequency sweep from 100 kHz to 0.01 Hz. All measurements were repeated twice with good reproducibility. The impedance spectra were analyzed with ZView software from Scribner Associates Inc (North Carolina, USA) with values of the goodness of fitted spectra corresponding to chi-squared <0.01 (square of the standard deviation between the original data and the calculated spectrum). The errors for the individual parameters of the equivalent electrical circuits were <5%. Additionally, immersion tests in 3.5 wt.% NaCl were performed for a duration of 168 hours to check the long term stability of the coatings and to characterise the corrosion products by XRD.

*Photoluminescence spectral measurements* were taken on a Horiba Jobin Yvon Fluorolog FL3-22 spectrofluorometer (PL, Edison, NJ, USA) at room temperature, with a 450W xenon lamp as the excitation light source. Obtained spectra were corrected for the spectral response of the measuring system and spectral distribution of excitation light source. Presented results are averaged over three samples for each current density in order to obtain representative PL spectra.

*Photocatalytic activity* of obtained coatings on Zn alloy substrate was determined by degrading methyl orange (MO) at room temperature. Samples were immersed into 10 mL of 8 mg/L aqueous MO solution and placed on a perforated holder with a magnetic stirrer underneath. Prior to irradiation, the solution and the catalyst were magnetically stirred in the dark for corresponding irradiation time to check coatings' adsorption properties. For irradiation HPR 125 Philips high vapour pressure mercury lamp (125 W) was placed 25 cm above the top surface of the solution. A fixed quantity of the MO solution was removed every 2 h to measure the absorption and then to determine concentration using UV-Vis spectrophotometer Thermo Scientific Evolution 220 (UV-Vis, Waltham, MA, USA). After each measurement probe solution was returned back to the photocatalytic reactor. Prior to photocatalysis measurements, MO solution was tested for photolysis in the absence of the photocatalyst in order to examine its stability. The lack of change in the MO concentration after 8 h of irradiation revealed that degradation was only due to the presence of photocatalyst.

## Results and Discussion

*Microstructure and phase composition*

The coatings are mainly composed of $ZnAl_2O_4$ and $ZnO$ with minor amounts of $AlPO_4$ and $Zn_2P_2O_7$. The coating formation starts with $ZnO$ at lower discharge energies and changes to a reactive formation of $ZnAl_2O_4$ when the discharge energy is increasing. Thus, the main differences in elemental and phase composition were found for the treatments done for only 3 minutes or at a current density of 75 mA cm$^{-2}$. However, as soon as the final voltage exceeds 300 V the main phase of the coating is $ZnAl_2O_4$. To see this influence on the distribution of the elements and correlate it with the phase distribution in the coating, elemental mapping (**Fig. 1**) was done for the cross sections of coatings prepared at the lowest current density of 75 mA cm$^{-2}$ (lower discharge energy) and a higher current density of 125 mA cm$^{-2}$ (higher discharge energy). At the lowest discharge energy (3 min, 75 mA cm$^{-2}$) $ZnO$ is the main coating phase, but surprisingly the $ZnO$ is obviously covered by an $AlPO_4$ surface layer, which may contain some $Zn_2P_2O_7$ as well. At the other conditions mixed oxides from $ZnO$ and $Al_2O_3$ are the main components as Zn and Al are jointly distributed over the coatings. As the Al source is the electrolyte and Zn comes from the substrate there is a trend that Al signals are stronger closer to the surface especially for longer treatment times (thicker coatings). Interestingly, phosphorus seems to be enriched closer to the interface, but it is not possible to assign it to Al ($AlPO_4$) or Zn ($Zn_2P_2O_7$) as the two phosphorus containing phases identified by XRD (**Fig. 2**). The results clearly show that with increasing discharge energy the amount (intensity) of the spinel peaks are increasing while some small remains of $ZnO$ are always visible which might be related to the interface barrier layer. The spinel phase should have positive effects on the wear and corrosion resistance [53-55], while the $ZnO$ should improve photoactivity of the specimen.

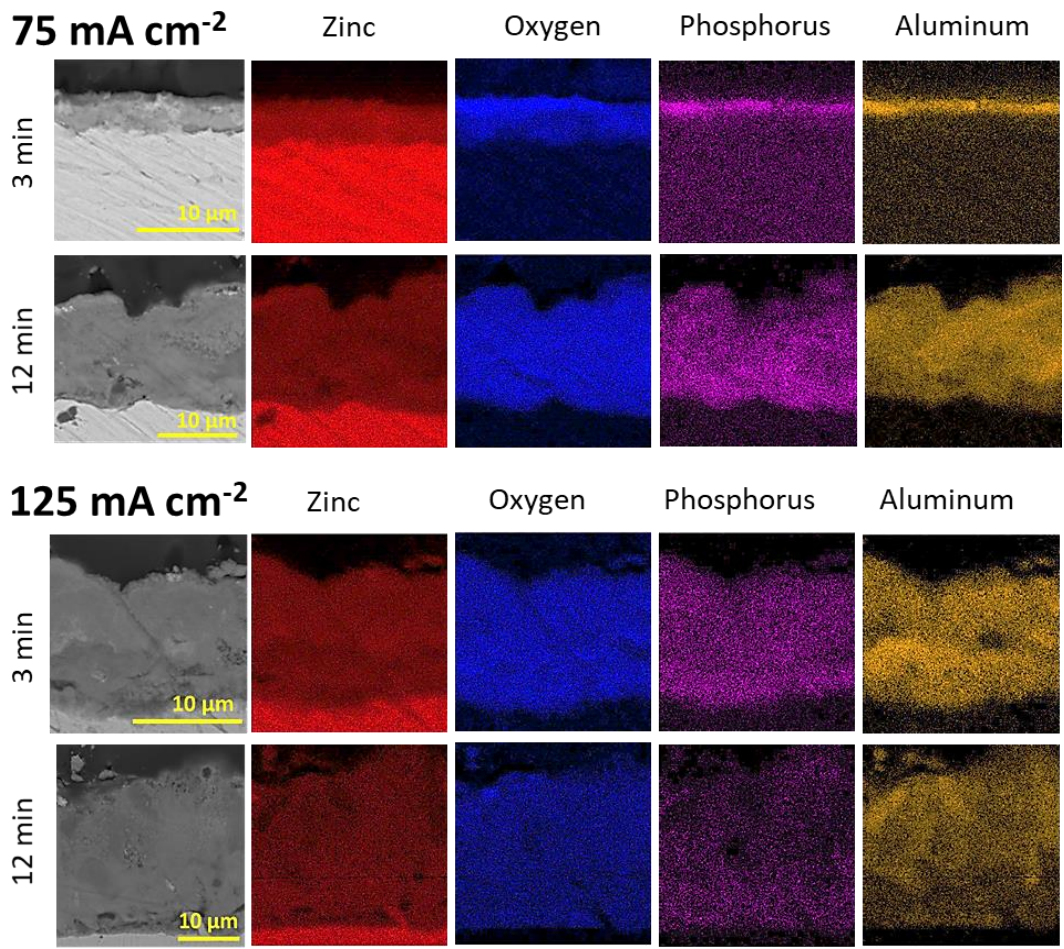

*Fig. 1: Elemental distribution through the PEO coatings obtained under 75 and 125 mA cm$^{-2}$ after 3 and 12 minutes treatment.*

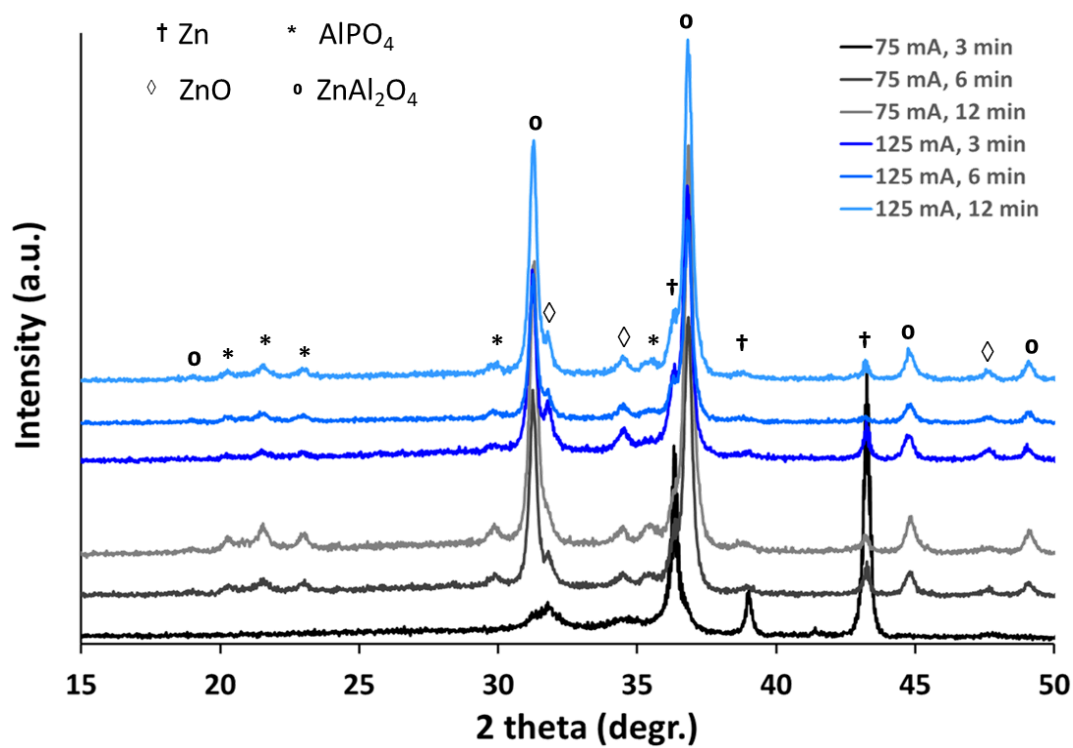

*Fig. 2: XRD patterns of PEO coatings obtained under 75 and 125 mA cm$^{-2}$ after 3, 6 and 12 minutes treatment.*

*Adhesion/cohesion behaviour*

**Table 1** summarizes the results of the pull-off tests. Adhesion/cohesion of the coatings was already a problem which was realized in previous work [51], but it was not quantified. The values are low, which might be related to the highly defective nature of the coatings, internal stresses, not sufficient sintering within the layer and easy crack growth along the pore band. A real trend is not visible in the data regardless if the current density or the treatment time was changed.

**Table 1: Results of pull-off tests, adhesion/cohesion strength measured with a dolly diameter of 20mm glued with Araldite 2011 to the specimen surface.**

| Current density, mA cm$^{-2}$ | Treatment time, min | σ, MPa | | | | Average σ, MPa |
|---|---|---|---|---|---|---|
| **75** | 3 min | 0,93 | 0,57 | 0,64 | 0,59 | **0,68 ± 0,12** |
| | 6 min | 1,12 | 1,16 | 0,91 | 1,26 | **1,12 ± 0,10** |
| | 12 min | 0,64 | 1,46 | 0,93 | 0,74 | **0,94 ± 0,26** |
| **100** | 3 min | 0,77 | 0,96 | 0,84 | 0,73 | **0,83 ± 0,08** |
| | 6 min | 0,63 | 0,84 | 0,74 | 0,83 | **0,76 ± 0,08** |
| | 12 min | 0,68 | 0,86 | 0,88 | 0,62 | **0,76 ± 0,11** |
| **125** | 3 min | 0,69 | 1,43 | 1,24 | 0,73 | **1,02 ± 0,31** |
| | 6 min | 1,44 | 0,73 | 0,48 | 0,98 | **0,91 ± 0,30** |
| | 12 min | 0,86 | 0,79 | 0,75 | 0,79 | **0,80 ± 0,03** |
| **150** | 3 min | 0,67 | 1,26 | 1,05 | 0,81 | **0,95 ± 0,21** |
| | 6 min | 0,69 | 0,76 | 1,22 | 0,71 | **0,85 ± 0,19** |
| | 12 min | 0,88 | 0,90 | 1,19 | | **0,99 ± 0,13** |

Optical and SEM observation of the specimen and dolly surfaces after the pull-off test do reveal a cohesive failure within the coating layer (**Fig. 3, 5** and **6** respectively). The crack has appeared inside the PEO layer, separating it in a way that coating remains on both sides with the outer layer staying fully on the dolly and the inner layer of the coating remains on the substrate.

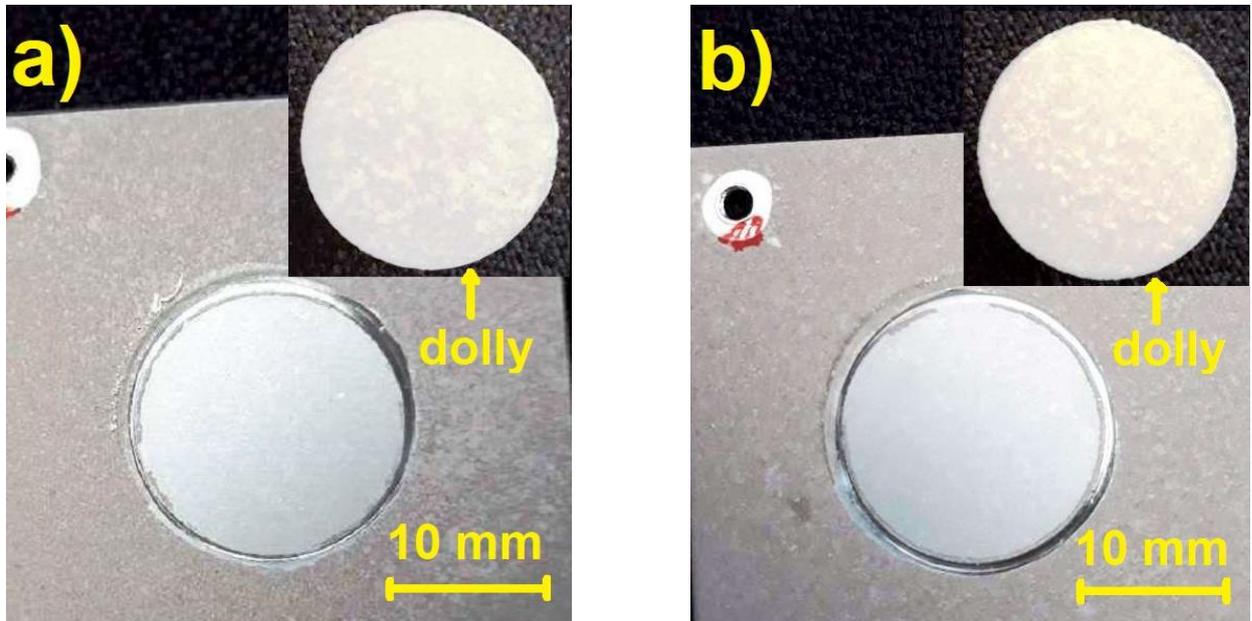

*Fig. 3: Two examples of the macroscopic appearance of the specimen and dolly surface side after pull-off test a) 12 min under 75 mA cm$^{-2}$ and b) 6 min under 100 mA cm$^{-2}$ respectively.*

To proof the visual observation, a larger area EDS analysis (full area analysis at magnification of 1kx) was performed on both sides and the results are summarised in **Fig. 4**. Looking especially at the oxygen, aluminium and zinc content it is most likely that more coating volume sticks on the dolly side as more zinc and less oxygen is detected on the substrate side, indicating excitation of substrate by the incoming electron beam. In contrast almost no change in the aluminium content is observed on the dolly side as well as a relatively uniform zinc and oxygen levels, indicating that the information is coming mainly from the coating and not from the Al dolly. The phosphorus content is higher for thinner layers on the dolly side which suggests that phosphorus is enriched in the interface between substrate and coating.

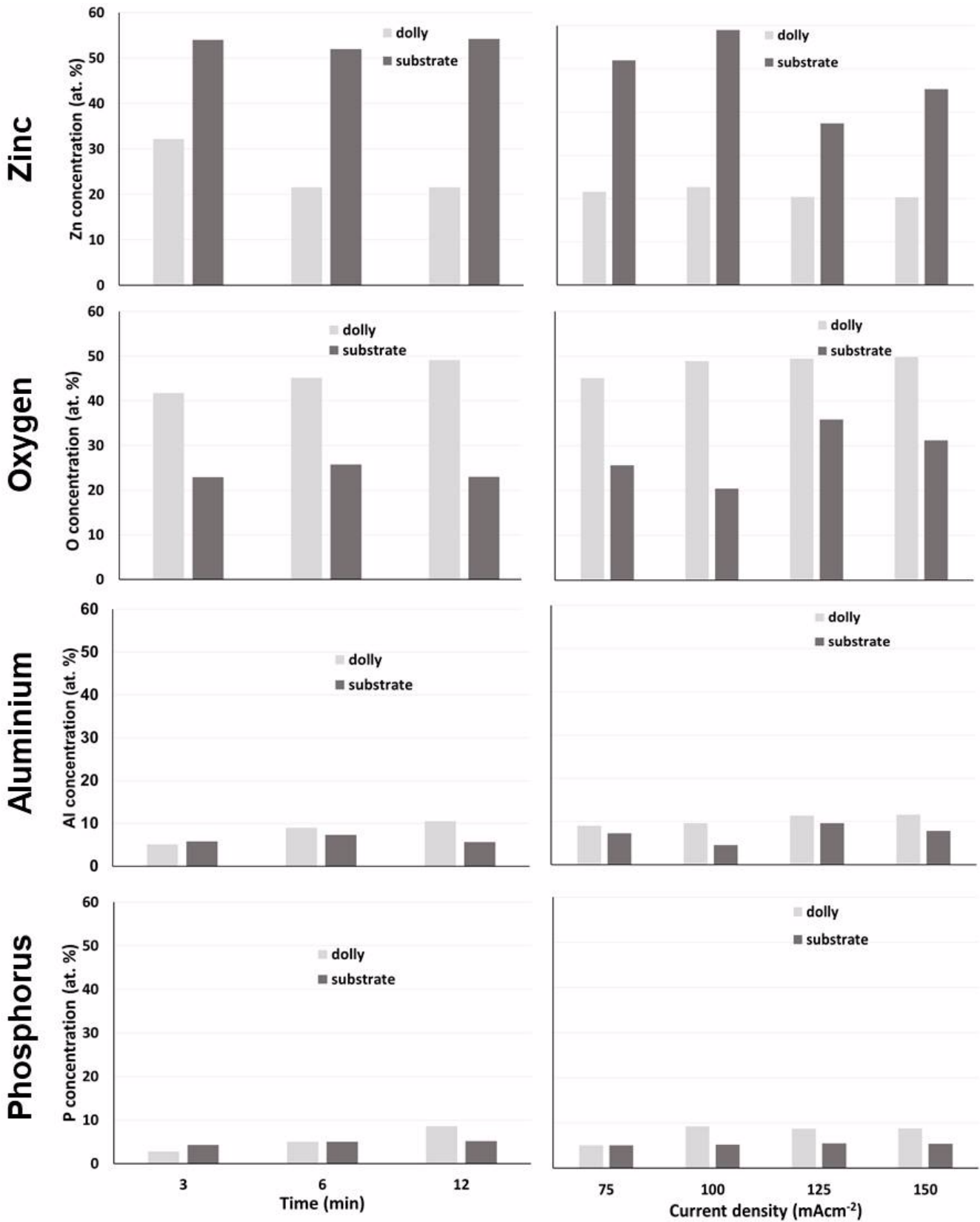

*Fig. 4: Composition (only main elements) of the various coating remains on the dolly and substrate sides after pull-off test (in at. %); left) 3, 6 and 12 min under 75 mA cm$^{-2}$ and right) 6 min under 75, 100, 125 and 150 mA cm$^{-2}$ respectively.*

**Figs. 5** and **6** show typical SEM micrographs of the fracture at dolly and substrate sides. There are coating remains on both sides which have a similar appearance. The coatings exhibit a large volume of internal porosity. Internal original surfaces with typical pores (solidified discharge channels) caused by discharges are visible within the coating. They are from discharge events inside of the coating e.g. between the interface barrier layer and the internal pore band. These internal surfaces appear to be rough and offer many starting points for crack formation. The cracked regions are also not densely sintered and do contain large amounts of small round pores (**Fig. 7**). The appearance of the coatings is more like a sponge, rather than a dense material. Only for the thinnest coating (3min, 75 mA cm$^{-2}$) the total volume of porosity is lower and the size of pores is smaller (**Fig. 5**), but a difference in cohesion strength is not noticeable.

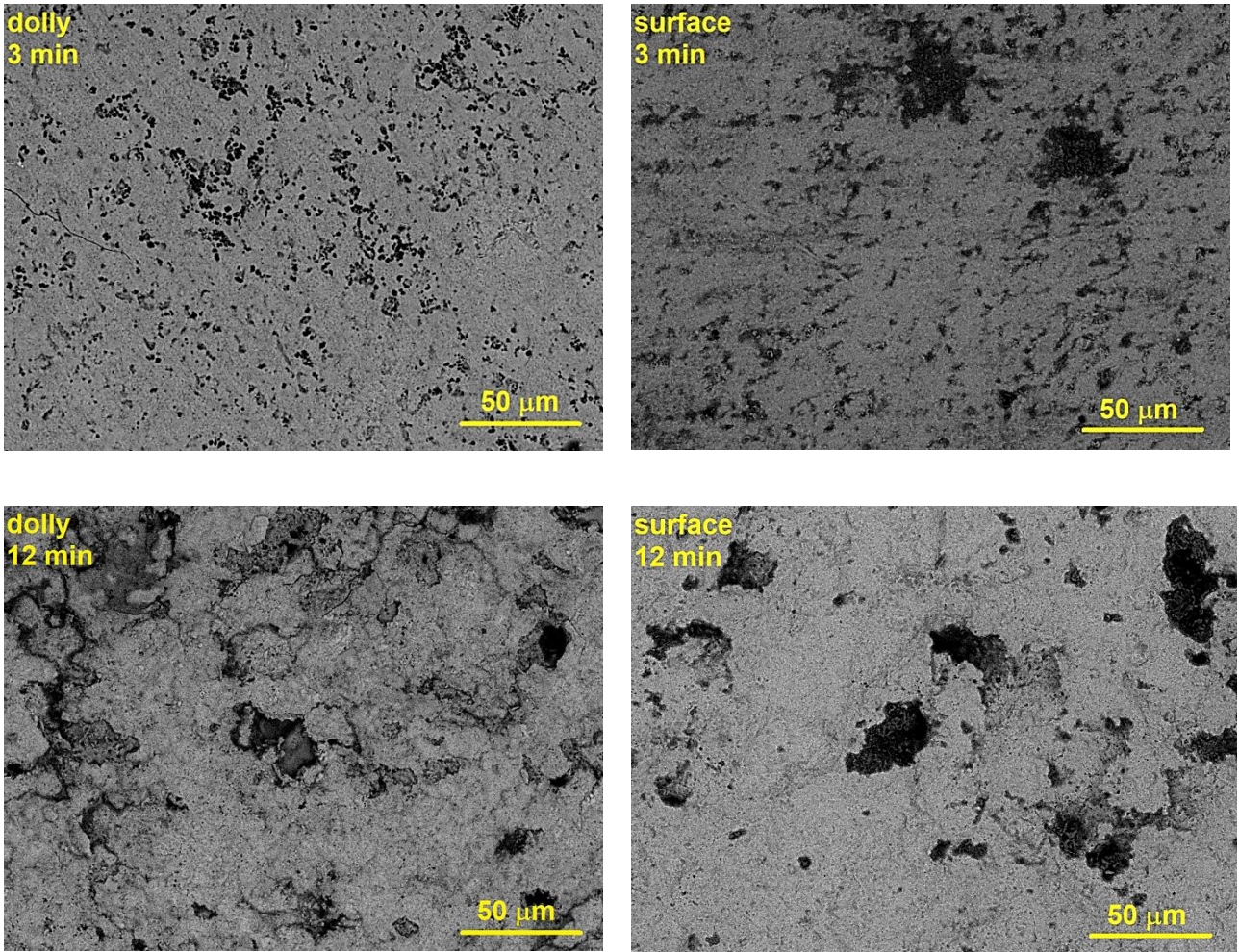

*Fig. 5: Example of the appearance of the fractures surfaces for specimens produced at 75 mA cm$^{-2}$ and two different treatment times; the left side shows the dolly and the right side the substrate*

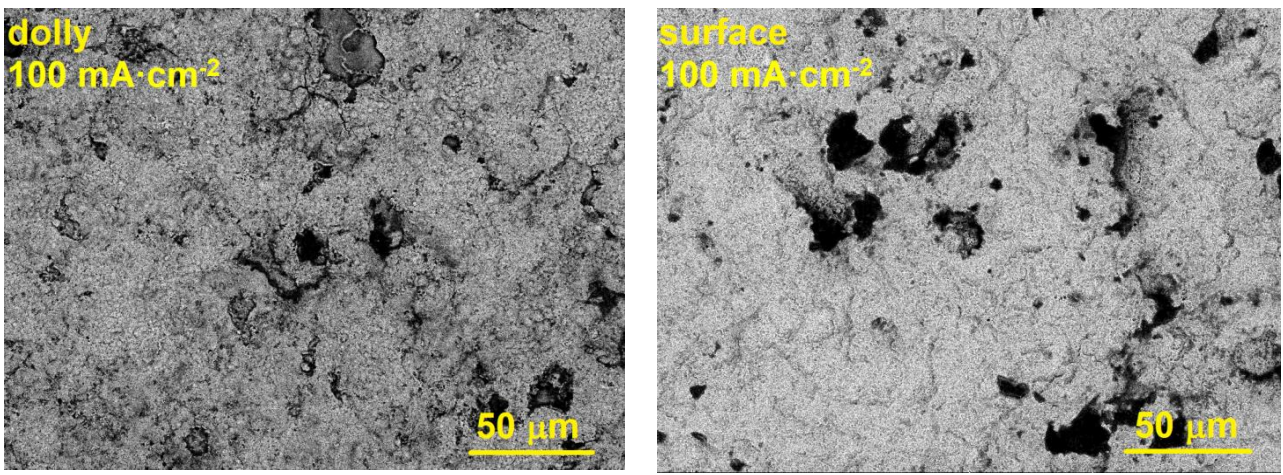

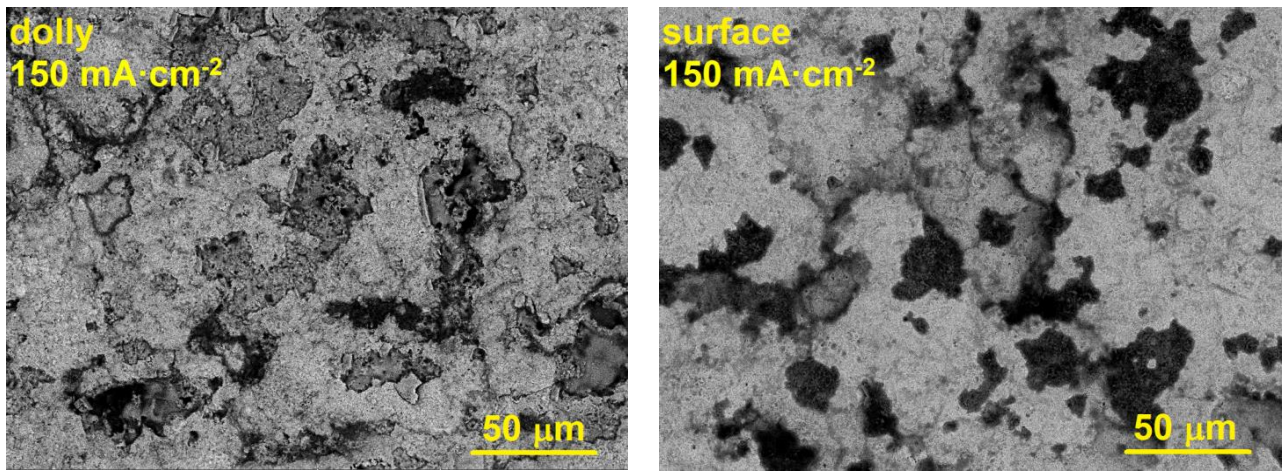

*Fig. 6: Examples of the appearance of the fractured surfaces for specimens produced at 6 min and two different current densities; the left side shows the dolly and the right side the substrate*

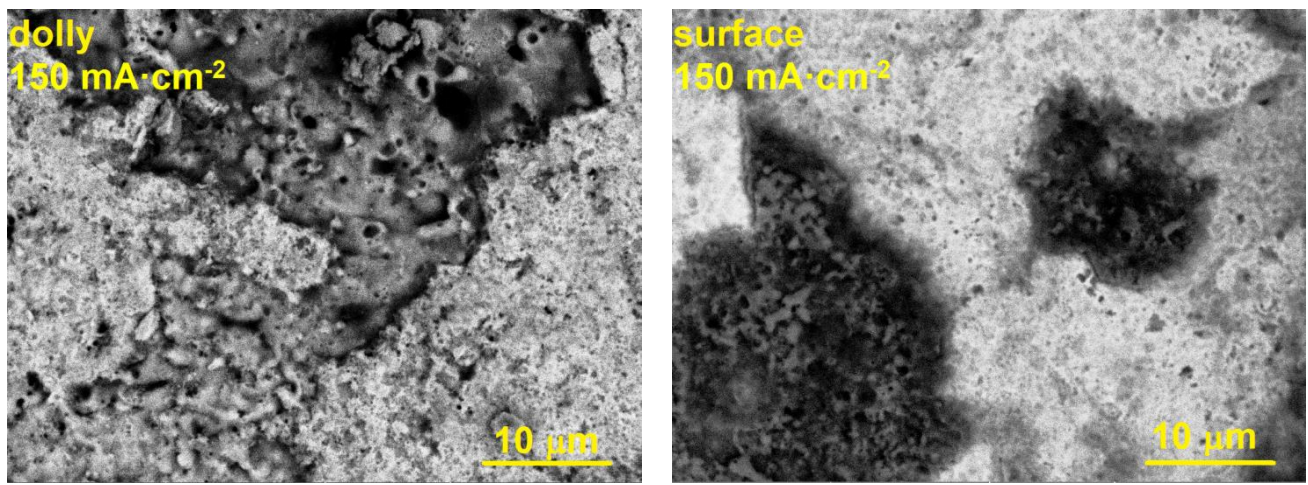

*Fig. 7: Fractured coatings at higher magnification revealing the sponge-like structure. The example shows the coating after 12 min treatment at 150 mA cm$^{-2}$*

*Wear behaviour*

To understand how the poor cohesion within the layers affects practical mechanical properties wear tests were performed. The wear tests were performed under decreasing load intervals (2 or 1 N load steps down to 1 N basic load) to determine the maximum load which the coating can withstand without failure. The coating failure was defined as total removal of the coating during the wear test. At 1 N load, the coatings formed at currents of 100-150 mA cm$^{-2}$ do not fail during the test and the substrate is not reached. Only in the case of the sample of 75 mA cm$^{-2}$; 3 min (when the coating is very thin) abrasion down to metal is observed immediately, which is obvious by the strongly oscillating friction force/friction coefficient around 0.75 +/- 0.1. When the coating remains intact, a relative stationary friction force is observed after the running-in period (**Fig 8a**). Looking at the friction coefficient it is obvious that due to the coatings the wear mechanism changed from more adhesive fretting dominated wear for the uncoated Zn substrate to a milder abrasive wear which affects also the ball in the case of the coatings (**Fig. 9**). Transfer of Zn to the steel ball and even sticking of coating flakes in the transferred Zn metal is visible in the case of a coating failure (**Fig. 9a**) and abrasive removal of material from the steel ball if the coating survives (**Fig. 9b**). Most of the coatings do survive 3 N loads, and only the thinner 3 min coatings at 75 and 100 mA cm$^{-2}$ and the 6 min 75 mA cm$^{-2}$ failed. At 5 N no reliable performance is found for the 75, 100 and 125 mA cm$^{-2}$ coatings anymore, even for the thickest coatings (**Fig. 8b,c**). Interestingly the performance of the coatings produced at 125 mA cm$^{-2}$ was worse than those formed under 100 mA cm$^{-2}$. Only the thickest two coatings produced at 150 mA cm$^{-2}$ do survive the wear test at 5 N load (**Fig. 8d**). All the results are summarised in **Tab. 2**. Please note that none of the coatings survived 6 N load.

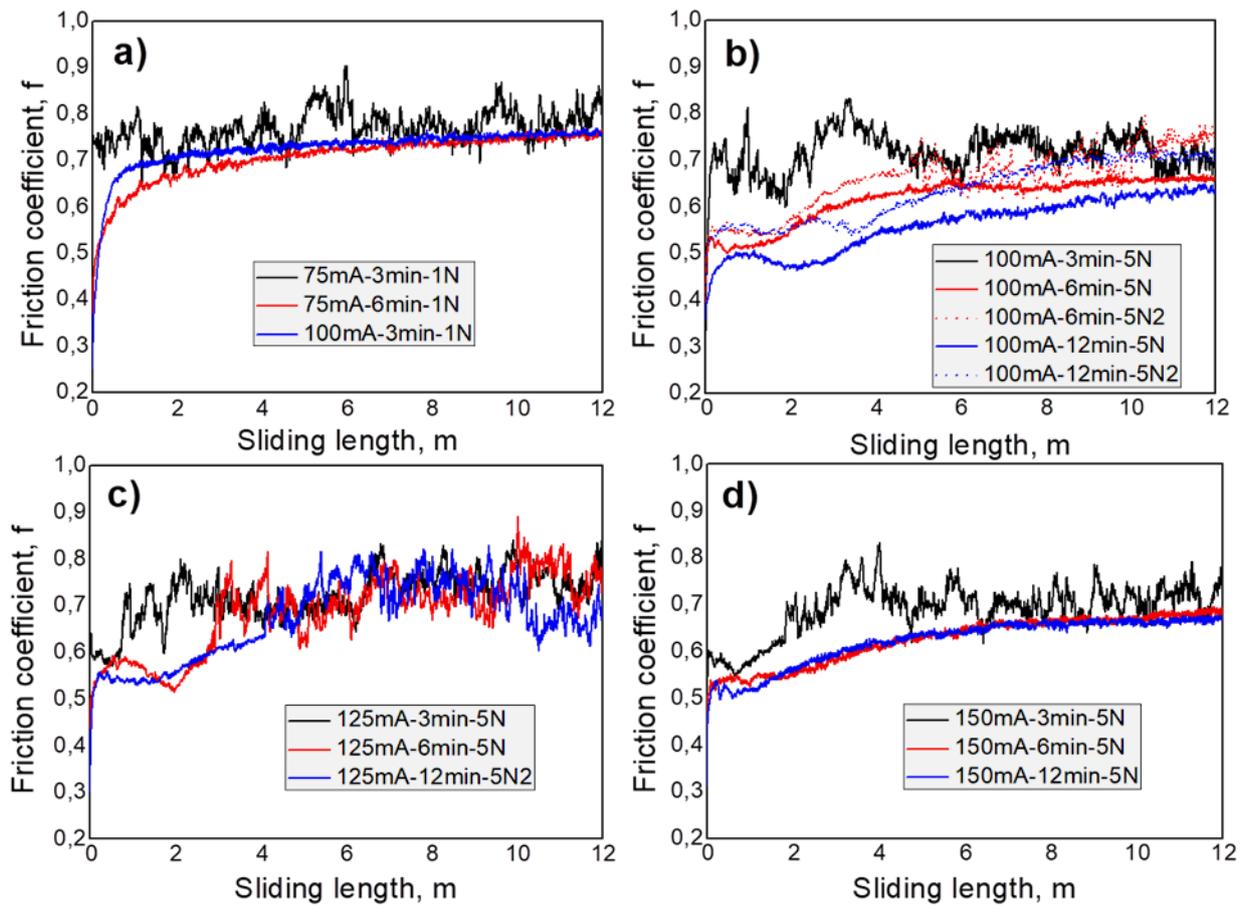

*Fig. 8: Examples of recorded friction coefficients against sliding distance for various coatings and loads (running against 100Cr6 steel ball) a) performance of thinnest coatings at 1 N load, b) performance of 100 mA cm$^{-2}$ coatings at 5 N, c) performance of 125 mA cm$^{-2}$ coatings at 5 N and d) performance of 150 mA cm$^{-2}$ coatings at 5 N (please note that 5N2 means the second repetition at 5 N load)*

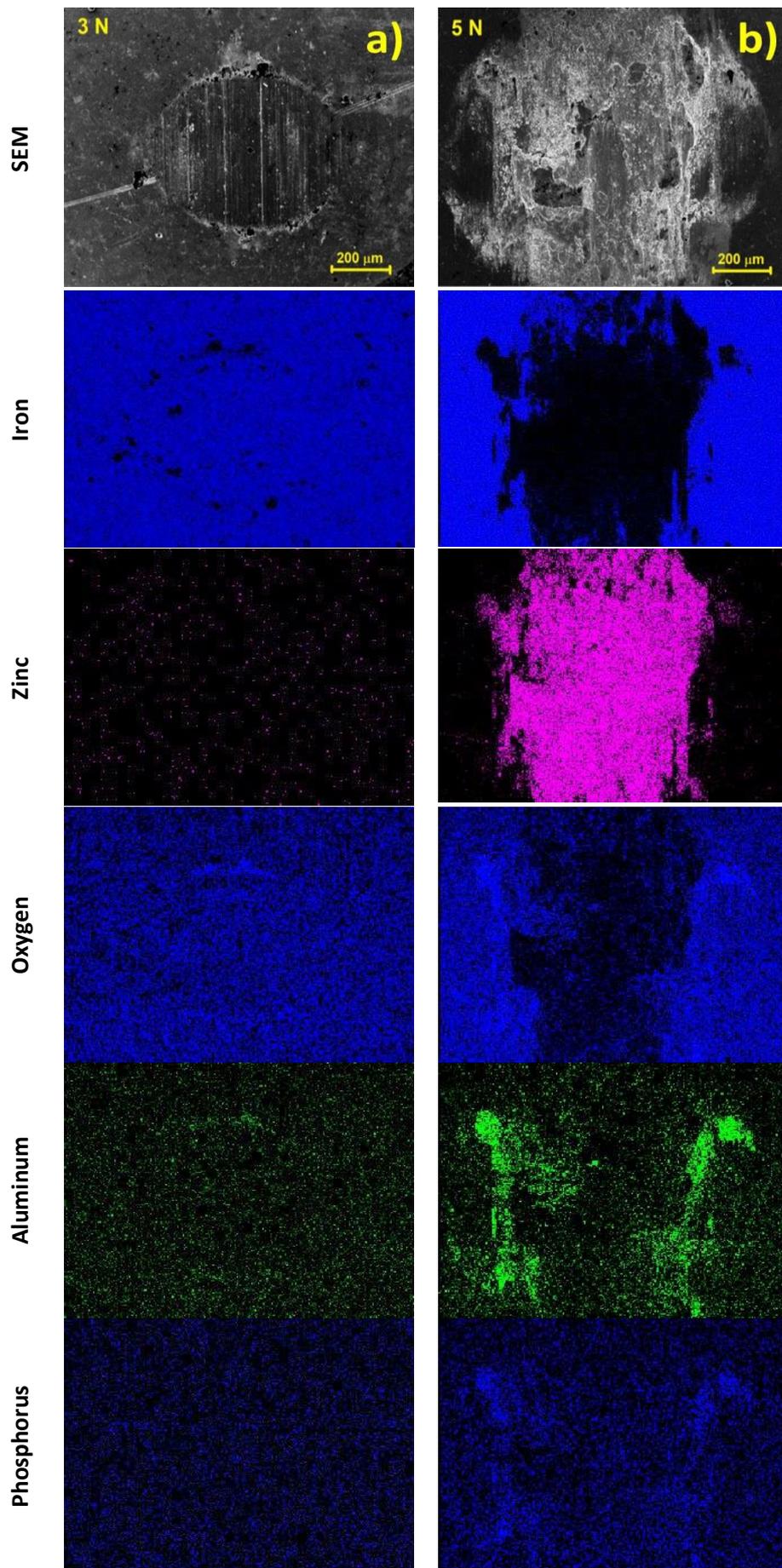

*Fig. 9: SEM micrograph and elemental distribution map of the surface of 100Cr6 steel ball after the wear test. The data is shown for a) 125 mA cm$^{-2}$_6min_3N – left column and b) 125 mA cm$^{-2}$_6min_5N – right column.*

Table 2: Overview of wear result of the investigated coatings against 100Cr6 steel ball

| Load: | 75 mA cm$^{-2}$ | | | 100 mA cm$^{-2}$ | | | 125 mA cm$^{-2}$ | | | 150 mA cm$^{-2}$ | | |
|---|---|---|---|---|---|---|---|---|---|---|---|---|
| | 3 min | 6 min | 12 min | 3 min | 6 min | 12 min | 3 min | 6 min | 12 min | 3 min | 6 min | 12 min |
| 5N | fail | fail | fail | fail | fail | OK | fail | fail | fail | fail | OK | OK |
| 3N | fail | fail | OK | fail | OK | OK | OK | OK | OK | OK | OK | OK |
| 1N | fail | OK | - | OK | - | - | - | - | - | - | - | - |

The SEM and the laser microscope observation of the wear tracks are shown in **Fig. 10**. There are only two conditions and two representative coatings after the wear tests shown. If the coating survives, the spongy structure of the coating is compacted and pressed into the soft Zn substrate. In the case of a coating failure the remains of the coatings are removed from the wear track and almost no oxidation of the naked Zn surface is observed. From the data obtained by the laser microscope the wear volume was calculated and displayed in **Fig. 11**.

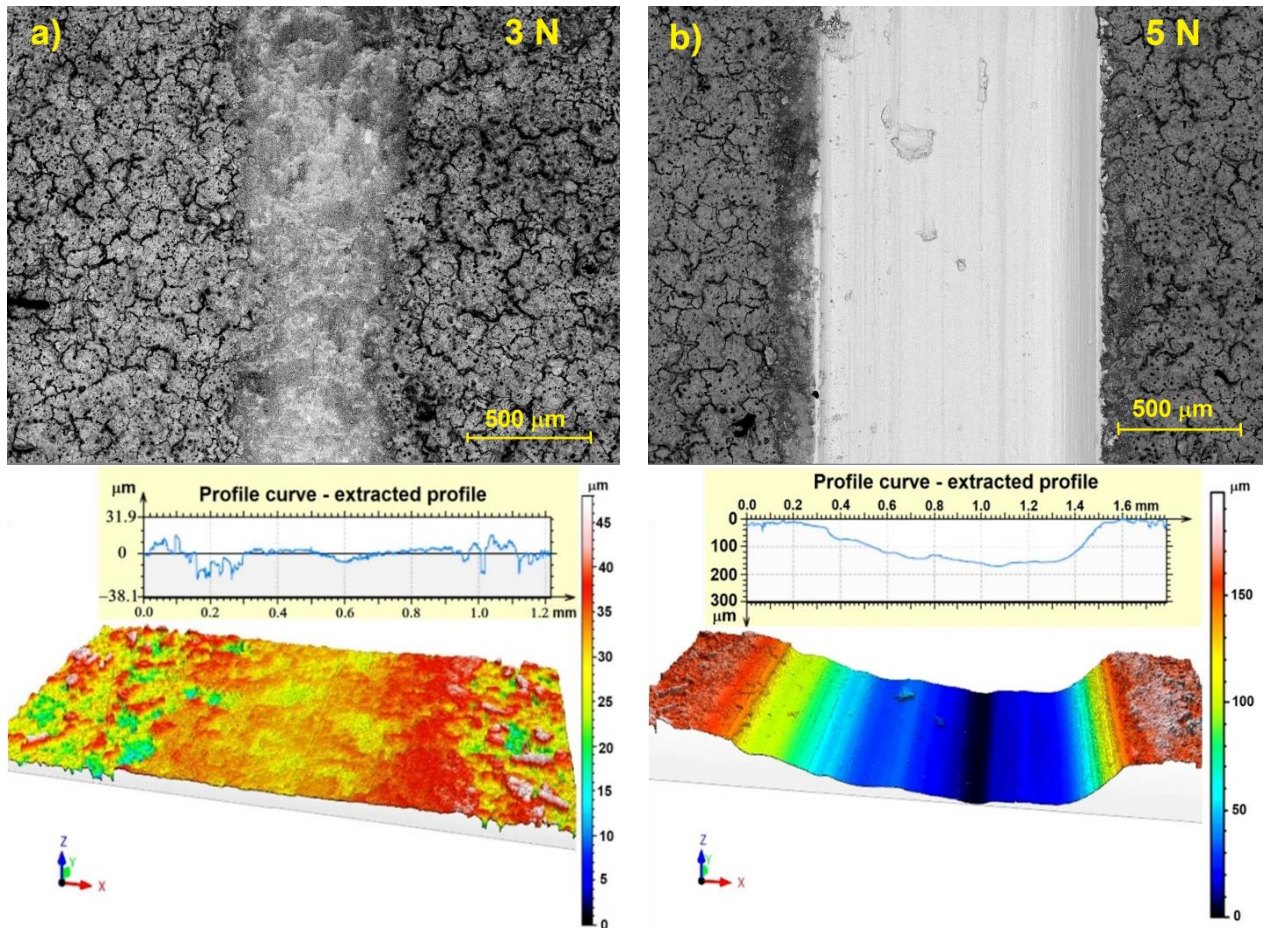

Fig. 10: SEM micrograph and LSM surface profile of a coating, which a) survives (125 mA cm$^{-2}$_6min_3N) and b) fails (125 mA cm$^{-2}$_6min_5N) during the wear test

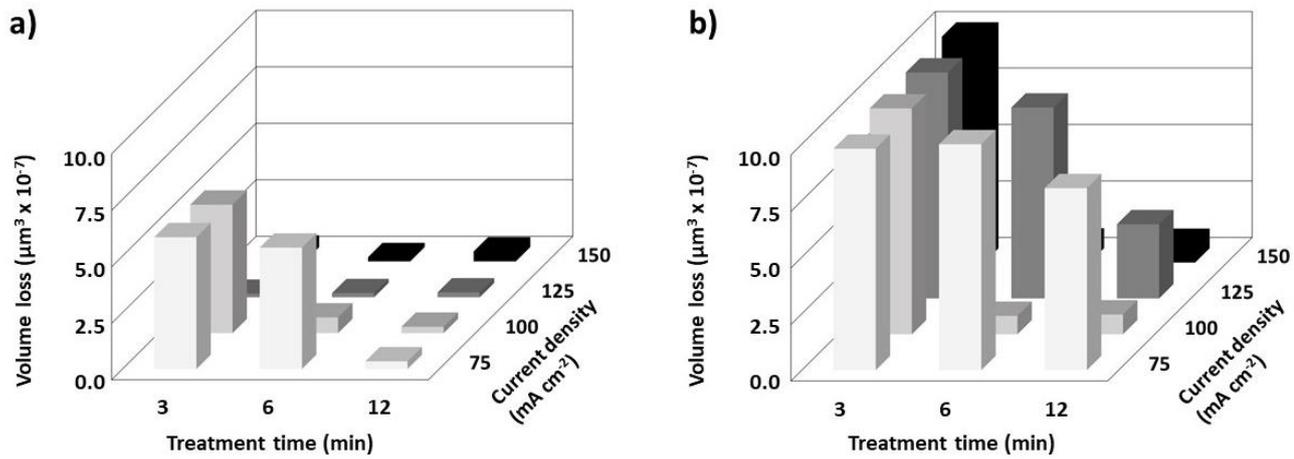

*Fig. 11: Volume loss of the substrate/coating systems at loads of a) 3 and b) 5 N after 12 m dry sliding against a 100Cr6 (6 mm diameter) steel ball.*

*Corrosion behaviour*

The corrosion properties normally depend strongly on the porosity of the PEO layers and the inner barrier layer. One can see how the sponge like PEO coating performs as corrosion protection for the Zn substrate. **Fig. 12** shows the Bode plots and phase angle evolution of the thicker coatings obtained after 12 min of PEO treatment at the different current densities. Spectra obtained on bare Zn were added as reference. Corrosion resistance of PEO coated Zn improves compared to bare Zn. It becomes higher with increasing current density as well as with immersion time of the obtained coating in NaCl solution. On the contrary, corrosion resistance of bare Zn corrosion is continuously decreasing while immersed in chloride solution.

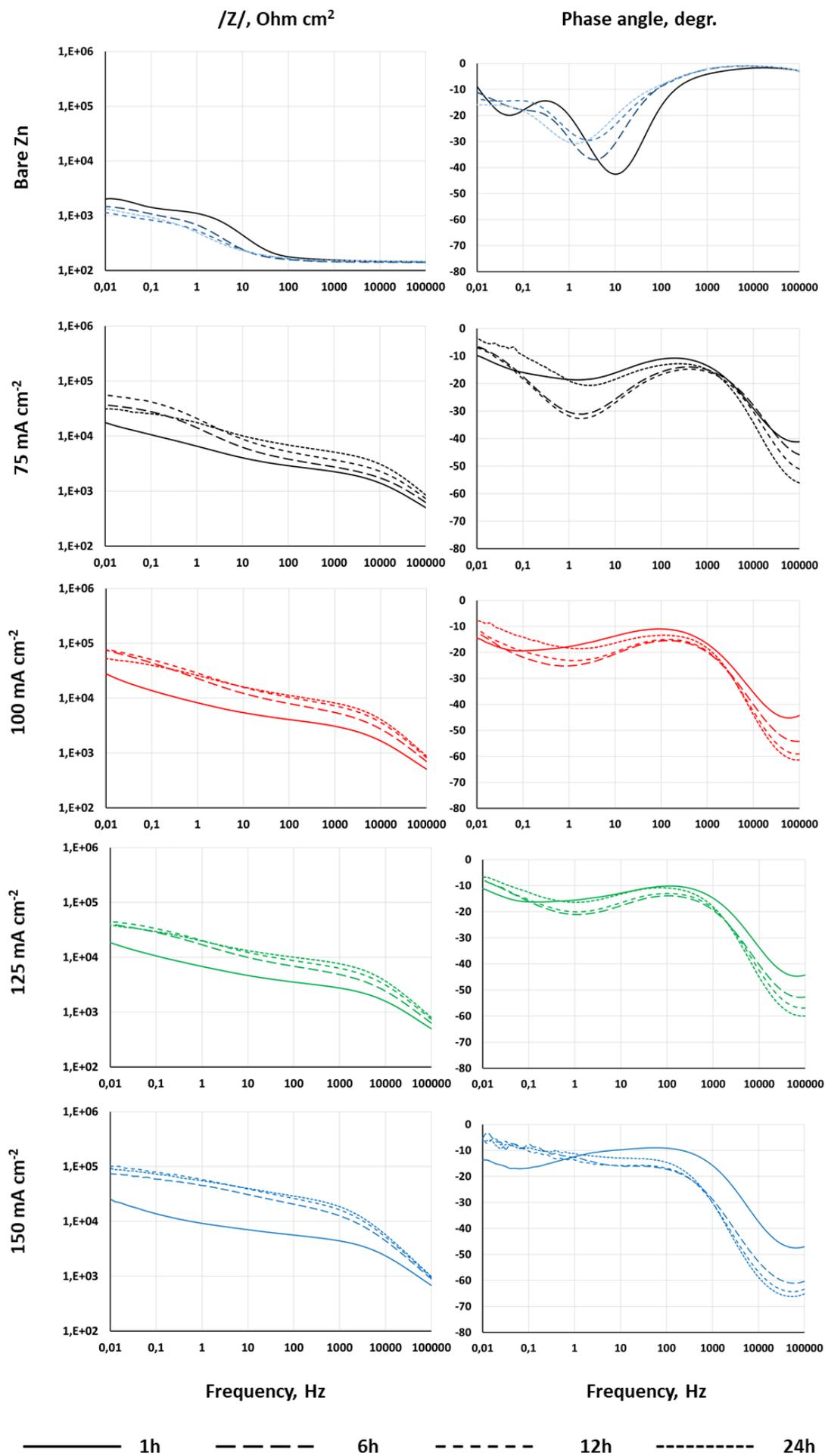

Fig. 12: *Bode plots of bare and PEO coated Zn (12 min treatment time) in 0.5% NaCl after 1, 6, 12 and 24 hours of immersion.*

The PEO coated Zn demonstrates two well defined relaxation processes. The high frequency time constant observed at about 1000-10000 Hz can be responsible for the barrier properties of the PEO layer. Most probably, the highly porous part of the coating is not contributing to the respective resistance and capacitance, since this layer is highly permeable for electrolyte, which shunts the porous layer. Therefore, the denser part of the coating next to the interface is the one, which is reflected at the high frequencies. The CPE is related to the capacitance of the interface layer, while resistance connected in parallel reflects resistance of electrolyte in its defects. The second relaxation process is evidenced as a suppressed time constant at about 1 Hz. This time constant can be attributed to the corrosion processes at the metal/electrolyte interface. The highly suppressed nature of the relaxation process can also be related with additional diffusion limitations through a porous PEO layer. However, for a sake of simplification a combination of CPE in parallel with a resistor was used here to fit the experimental spectra. In this case CPE can be ascribed to the double layer capacitance, while resistor corresponds to the polarization resistance.

**Fig. 13** shows the equivalent circuit used to fit the experimental data and inserted is an example of the fitting for a typical coating. Good fits of the experimental data were obtained with chi-squared values in the range of 0.001–0.0001. $R_s$ corresponds to the solution resistance. $CPE_{PEO}$ and $R_{PEO}$ are associated with the capacitive and resistive response both of the porous and barrier part of the coating. Its outer part includes different morphologies of defects (through-going discharge channels, cracks and pores). The corrosion processes were described by the double layer capacitance on the electrolyte/metal interface ($CPE_{dl}$) and the polarization resistance ($R_{polar}$), which is the Faradaic charge transfer resistance related to electrochemical reactions in the same electrolyte/metal interface region. Constant Phase Element (CPE) was used instead of capacitances in order to account for the non-ideal behaviour of the system.

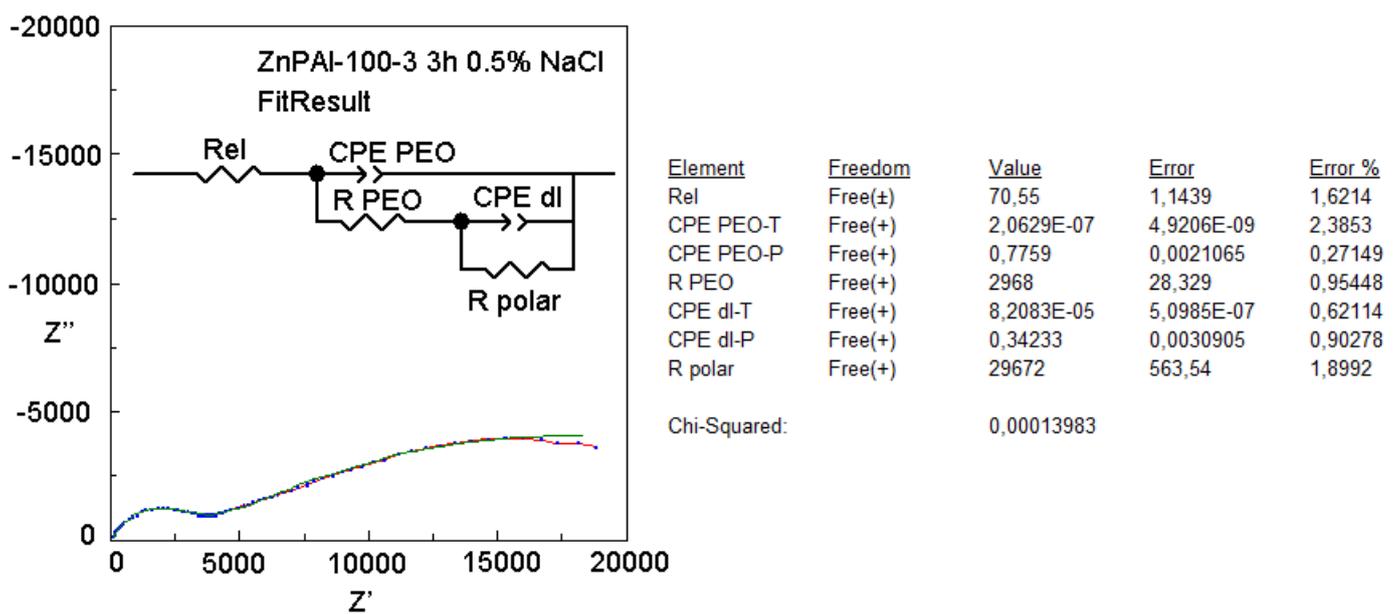

*Fig. 13: Equivalent circuit used to fit the experimental data, inserted in an example of the fitting of PEO coating (produced at 100 mA cm$^{-2}$ and 3 min) after 3h conditioning in 0.5 wt.% NaCl solution*

The PEO layer obtained after 3 min at 75 *mA cm$^{-2}$* has the lowest resistance (**Fig. 14**) and decreases at first. This is due to the destruction of the barrier layer after fast electrolyte penetration through the thin porous layer. However, for the film grown at 75 *mA cm$^{-2}$* the PEO resistance evaluation after 3h immersion in NaCl solution is uncertain due to its low value. If current density is up to 100 *mA cm$^{-2}$* then the obtained coating has the highest protection properties, which do not change much after 1 day immersion. Possible reason is that the film has less defects so it takes longer time for the electrolyte to migrate through it or the barrier layer at the interface is denser. Almost the same decrease of PEO layer resistance comparing with the film obtained at 100 *mA cm$^{-2}$* takes place if the process is performed at current density values of 125 and 150 *mA cm$^{-2}$*. Higher porosity of PEO and barrier layer for the latter films cause its lower total overall resistance. After 1 day immersion, the starting resistance of coatings grown at 125 and 150 *mA cm$^{-2}$* becomes higher coming closer to that grown at 100 mA/cm$^2$ and there is a positive effect visible with increasing treatment time/layer thickness on the time-dependence of the resistance during immersion (**Fig. 14**). PEO resistance depends mainly on its porosity, so sealing pores with corrosion products leads to the increase of PEO resistance with immersion time. PEO layer resistance values evaluated for the films produced at current densities between 100 to 150 *mA cm$^{-2}$* increase with immersion time (**Fig. 14**) and the best performance was observed for the longest treatment time and the highest current density. Capacitance is less for the films grown at higher current values as in latter case more thick coatings are formed. Decrease of capacitance with immersion time is possibly to the substitution of electrolyte solution in pores by corrosion products having less dielectric constant. One can estimate maximal coating thickness is about 1.25 µm taking into account that its capacitance exceeds 6·10$^{-5}$ F/m$^2$ and assuming that its dielectric constants is less than 8.5 which is its value for ZnO [56, 57]. The calculated value of coating thickness could be attributed to its inner barrier layer.

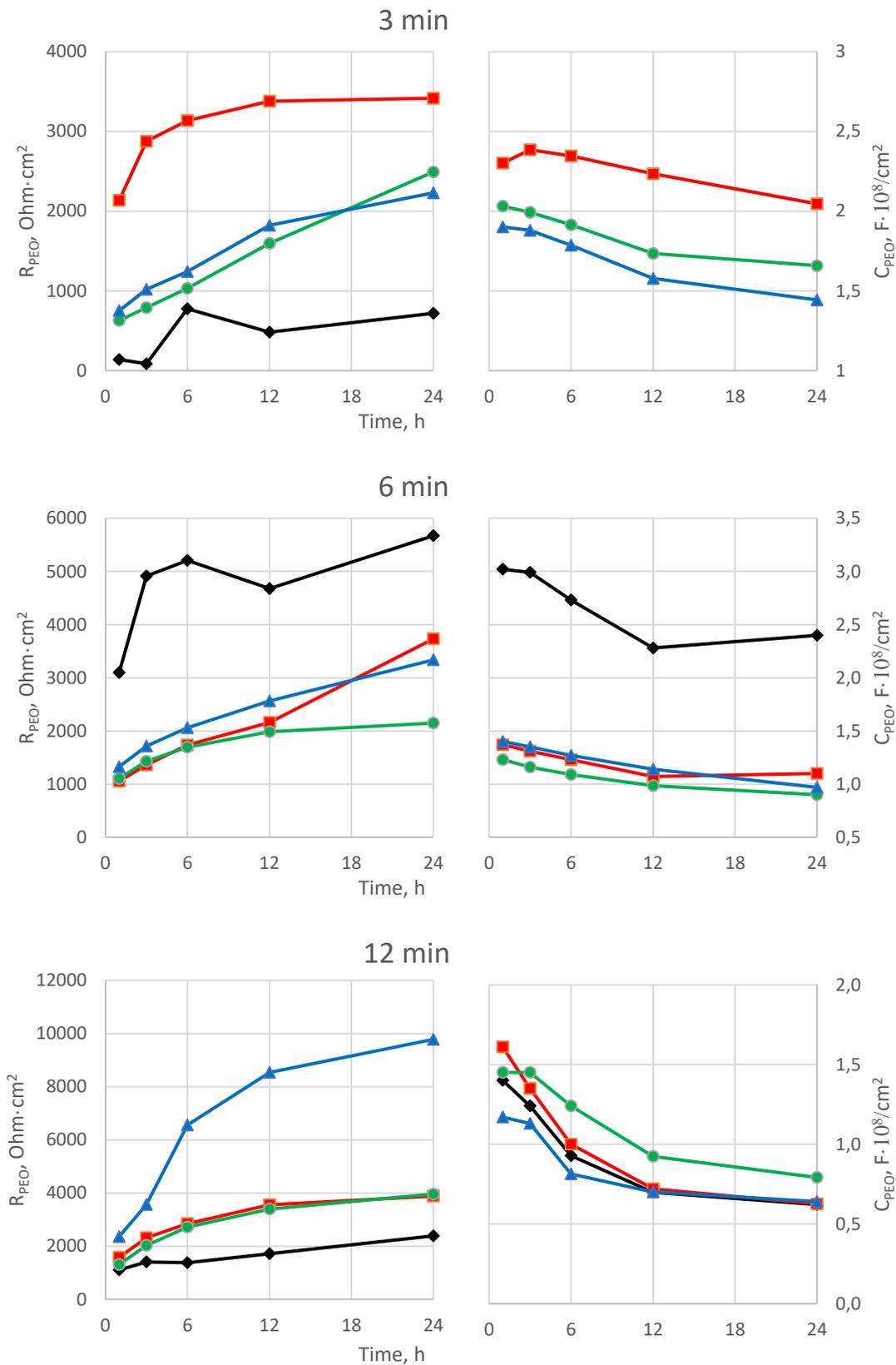

*Fig. 14. Evolution of PEO layer resistance (a) and capacitance (b) during 1 day of immersion in 0.5 wt.% NaCl. PEO layers on Zn grown for 3, 6 and 12 minutes at current densities of 75 (black) 100 (red), 125 (green) and 150 mAcm$^{-2}$ (blue).*

A long term stability test of 168 hours immersion into 3.5 wt.% NaCl solution was successfully passed for all specimens. Optically, no visible changes of the surfaces were detected. The XRD diffraction pattern before and after corrosion for the two types of coating compositions are shown in **Fig. 15**. A short time low current density version which is mainly composed of ZnO and a long time low current density version which is mainly composed of spinel were selected as representatives. It is obvious that the coatings are quite stable showing no dissolution of phases and only slightly higher intensity peaks of ZnO. Please note that ZnO and $Zn(OH)_2$ are having the same peak positions, thus it might be likely that the only corrosion product is $Zn(OH)_2$ under the laboratory test conditions.

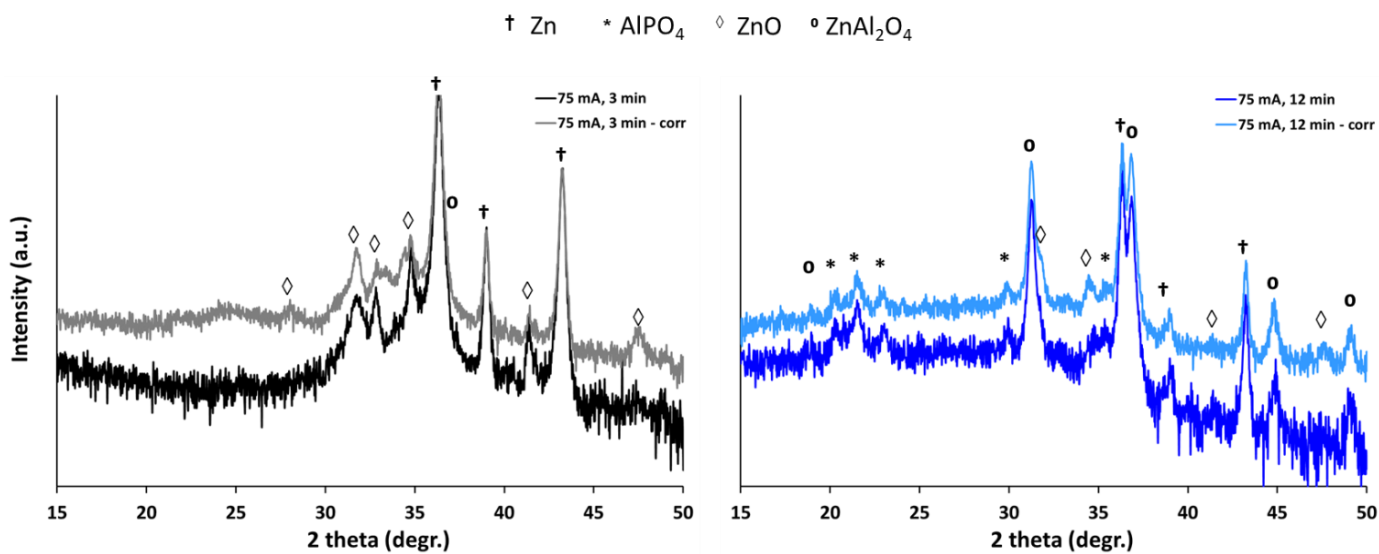

*Fig. 15. XRD patterns of PEO treated Zn specimens before and after 168 h immersion in 3.5% NaCl solution (75 mA cm$^{-2}$ treatments for 3 and 12 min are selected as an example)*

*Photoluminescent properties*

Photoluminescence emission spectra of coatings obtained on Zn alloy by PEO are presented in **Fig. 16**. The emission band is dominantly present in the visible range of the spectrum and its intensity increases with PEO time for each current density and with increasing current density for the same PEO processing time. Broad and asymmetric emission photoluminescence spectra indicate the presence of multiple emission centres emitting simultaneously. It can be observed that the shape and the position of photoluminescent maximum changes with increasing PEO time and/or current density. This change can be related to the formation of spinel zinc-aluminate phase ($ZnAl_2O_4$) on the expense of ZnO phase with increasing PEO processing time and/or current density [51]. For low current density and short PEO processing time (blue line in **Fig. 16a**) photoluminescence maximum is centred at about 500 nm corresponding to ZnO photoluminescence emission maximum [53], while with increasing PEO time and current density maximum features blue shift towards photoluminescent spectrum of $ZnAl_2O_4$ [58]. It is also worth mentioning that the absence of ultraviolet photoluminescence band inherent to ZnO (blue line **Fig. 16a**) can be related to rather poor crystallization of obtained oxide coatings [59].

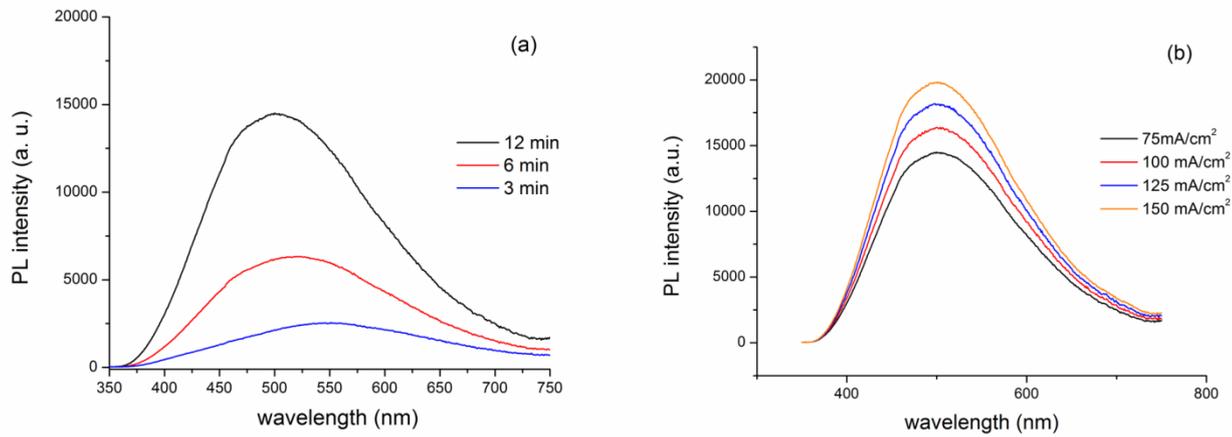

*Fig. 16: a) Emission PL spectra of coatings processed at 75 mA cm$^{-2}$ for various PEO processing time; b) Emission PL spectra obtained after 12 min of PEO with varying current density.*

Observed increase in photoluminescence intensity can be related to an increase in the number of defects (mostly oxygen vacancies) created in the oxide coating [60]. It has already been reported for this system that increasing discharge energy results in oxygen deficiency and that only for low current density and shortest PEO processing time the amount of oxygen detected in the near surface region can balance detected metal and phosphorus atoms [51].

*Photocatalytic properties*

Finally, the photocatalytic activity is of interest for possible applications such as waste water cleaning. Oxide coatings obtained by PEO on zinc alloy in phosphate-aluminate electrolyte are mainly composed by ZnO and ZnAl$_2$O$_4$ with smaller amounts of AlPO$_4$ and Zn$_2$P$_2$O$_7$, depending on the processing conditions [51]. A simple literature survey shows that these compounds have band gaps in the UV region of the electromagnetic radiation spectrum [61-64]. Therefore, it is not possible to use visible light for photocatalytic decomposition and one has to stay limited to UV light source for this purpose. Prior to photocatalytic testing, adsorption testing has been performed, i.e., PEO coatings have been placed on the perforated holder in stirred MO solution and kept in the dark for 8 h, checking MO concentration in 2 h periods. Adsorption and photocatalysis results for selected samples are presented in **Fig. 17a and 17b, respectively**.

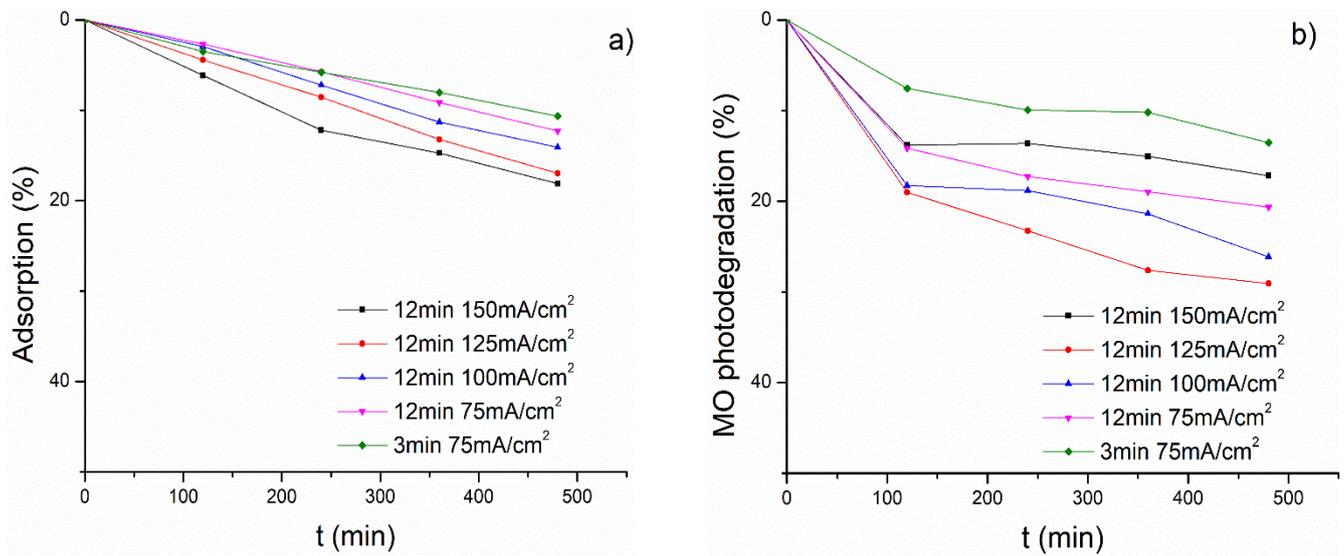

*Fig. 17: a) MO adsorption and b) photocatalytic decomposition of MO of oxide coatings obtained by PEO of zinc alloy in phosphate-aluminate electrolyte*

Results of adsorption measurements are presented in **Fig. 17a**. One can observe that adsorption increases linearly with time for all investigated samples. The sample processed for 3 minutes with current density of 75 mA/cm$^2$ showed the lowest adsorption of MO. Increase in processing time and current density results in increase of adsorption, thus the sample processed for 12 minutes with current density of 150 mA/cm$^2$ showed the highest adsorption of MO. These findings are in accordance with SEM results revelling the smallest total volume porosity and pore size for the thinnest coating (3min, 75 mA cm$^{-2}$). On the contrary, sponge-like structure and larger pores of PEO coatings facilitate adsorption.

The photocatalytic activity of PEO coatings under investigation, shown in **Fig. 17b**, is rather low. The lowest photocatalytic activity is observed for the sample processed for 3 minutes under 75 mA cm$^{-2}$ current density, although it is mostly composed by ZnO, which has the lowest bandgap value (3.2 eV) compared to ZnAl$_2$O$_4$ (3.9 eV), AlPO$_4$ (8.9 eV), and Zn$_2$P$_2$O$_7$ (3.6 eV). Only slightly higher photocatalytic activity has been observed for the sample processed for 12 minutes with current density of 150 mA cm$^{-2}$. Observed difference in photocatalytic activities of these two coatings cannot be explained only by difference in their porosity. The amount of ZnO for these two coatings is also very different and it is significantly higher for sample processed for 3 minutes under 75 mA cm$^{-2}$ current density. However, XRD [41] and photoluminescence results show poor crystallinity of ZnO in the later coating (3min, 75 mA cm$^{-2}$) which, together with its low porosity, results in the lowest photocatalytic activity.

In the case of other investigated samples photocatalytic activity can most likely be attributed to higher surface area accessible for photocatalytic decomposition and higher amount of ZnO. Indeed, the samples processed for 12 min with 100 and 125 mA cm$^{-2}$ current density contain the highest amount of ZnO [41] and exhibit the highest photocatalytic activity.

**Discussion**

In our previous work the focus was on coating and phase formation of the PEO coatings forming on Zn in an alkaline aluminate/phosphate based electrolyte [51]. However, a problem of coating adhesion/cohesion was already identified during PEO processing (which was addressed by a kind of repair processing) and the subsequent microstructural characterisation. The repair process appeared to work reasonably well with the optically best looking surface obtained after 12 min treatment at 150 mA cm$^{-2}$. In the current study we were looking in more details into the adhesion/cohesion properties of the layers and how they are affecting other properties such as wear and corrosion resistance, but also the photo-activity of the layers was of interest.

The pull-off tests revealed that there is no problem with the adhesion of the layer to the substrate, at least the bonding forces to the substrate are stronger than the internal bonds within the coating. The cohesion within the layer is relatively poor and forces of 1 MPa are sufficient to separate the coating. As one would expect, the crack/separation follows the large internal porosity, which was identified in the cross sections as a pore band closer to the coating/substrate interface. How extended and interconnected those large pores are, became quite obvious when looking at the fractured surfaces. In most of the cases around 50% of the total fractured area could be related to previous internal surfaces within the coating without any bonding to its opposite side. However, in addition to the already expected extended pore band, remarkable nano/micro porosity of the remaining coating was found. Obviously, those pores were closed during the mechanical grinding of the specimens for cross section preparation and not visible in the previous study. On the fractured surface they are clearly visible and the surface/coating looks like a sponge rather than a solid material (**Fig. 7**). However, those pores seem to be gas inclusions and are round in shape, thus a crack should not move that easily through those regions and may not be the only reason for the low bonding strength. It is possible that sintering of the layer was not that successful explaining the low bonding strength and ease to separate the coating. To form the spinel phase two oxides of the system needs to react:

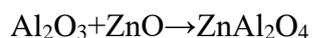
$Al_2O_3 + ZnO \rightarrow ZnAl_2O_4$

However the melting temperatures of those two oxides are quite high ($T_M$ (ZnO) =1975 °C and $T_M$ (Al$_2$O$_3$) =2072 °C) and the temperature range is close to the range of reported phases which were melted by the PEO process [65-67]. The energy was sufficient to form the phase, but not high enough to melt and densify larger volumes of the coating. The outcome is a porous coating with a mixture of large irregular pores of 50 to 100 µm in size, typical PEO pores from the discharge channels and gas inclusions up to a few micrometres in size and nano-pores with sizes less than a micrometre.

From this point of view it is interesting to look on the performance of the coatings as wear and especially corrosion depends strongly on the porosity and stability of the layer. Improvements compared to the untreated substrate are found for all studied properties. The wear resistance, which might be related to the fact that the coatings are more sensitive to tensile forces as applied during the pull-off tests rather than to more compressive forces during the wear test. Compared to other PEO coatings with similar thickness the load bearing capacity is lower [68-72], but still acceptable. Under the applied loads the coatings are compacted and pressed into the substrate, but they are not failing immediately. This might be related to the fact that under the pore band region which is responsible for the failure under tensile load still lays a relatively intact less defective layer of a few micrometres in thickness (estimated from cross sections [51] and electron excitation depth during EDS analysis of fractured surfaces). This layer can still support the top layer in the case that the bigger pores are collapsing under the compressive loads. Another positive effect of the coatings is a change of the wear mechanism. The substrate suffers from strong adhesive wear between the soft zinc and the hard steel ball. Thus, an unlubricated test between Zn and steel cannot be performed, because during wear testing the safety limit of the wear tester regarding the friction force is quickly reached and it turns off automatically. With the coating applied, this is not happening anymore, but in the case of a coating failure the strong oscillation of the friction force is still visible, giving evidence of the strong adhesion between the two metals. The friction coefficient of the coating and the steel ball is not much lower, but there are no large oscillations in the friction force demonstrating the change to a mild abrasive wear which is not only restricted to the substrate but also occurs on the ball.

Porosity is also affecting the corrosion resistance. The corrosion behaviour monitored by EIS shows a typical behaviour of a "sponge-like" material. The electrolyte is easily penetrating the open porosity of the outer layer reaching the interface towards the barrier layer. Within the coating the exchange of electrolyte is limited and mainly depends on diffusion. In contact with the electrolyte the less stable phases of the outer porous layer and of the barrier layer are dissolving and corrosion products can form. Apart from Zn ions from the substrate and the coating, also the remains of the PEO treatment solution (containing phosphates and hydroxides) in the internal porosity of the film and the corrosive solution with chloride ions and dissolved $CO_2$ (carbonate ions) need to be considered in formation of possible corrosion products. In general, the corrosion of Zn in different environments is a complex process and a variety of possible products may form [73, 74]. Most of them e.g. $ZnO$, $Zn(OH)_2$, $Zn_3(PO_4)_2 \cdot 4H_2O$, $ZnCl_2 \cdot 4Zn(OH)_2$ and $Zn_4CO_3(OH)_6 \cdot H_2O$ have a low solubility in water and will form solid precipitates and as a consequence the open porosity of the PEO layer is filled. This was observed by an increasing coating resistance during the EIS measurements. However, the long term immersion test suggests that the conditions during the test in the laboratory are not that complex and the only detectable corrosion products are $ZnO$ or $Zn(OH)_2$ and the coating phases are quite stable without detectable changes. From this point of view the porosity is not as critical as it contributes to a stabilisation of the layer offering a kind of self-healing ability. Furthermore, the porosity can be used to functionalise the coating for example if it is used as storage for inhibitors, drugs or other functionalised compounds such as nanocontainers. Post sealing of the coating with a polymer coating/paint may also contribute to improved cohesion in such a composite coating as defects are sealed and load is distributed over two interconnected materials. These are possible future activities to utilise the sponge-like surface structure.

The rough and porous surface may also contribute to the observed photoluminescence, photoactivity and adsorption of the spinel dominated coatings. Clearly, photoluminescence and adsorption rise with higher applied voltage and prolonged PEO processing times, suggesting that rougher and/or more porous coatings possess higher number of luminescence centres and/or active adsorption sites. Such complex behaviour of obtained coatings is also related to their change in phase composition due to the different processing parameters. In the coatings obtained with low voltages and short PEO times the dominant phase is ZnO, but PL and XRD show its rather poor crystallinity. Furthermore, the elemental mapping results indicate that the ZnO layer is covered by $AlPO_4$, which may explain the poor photoactivity of the coating which should have the highest amount of ZnO (3 min, 75 mA cm$^{-2}$). Absorption bandgap of ZnO matches the wavelength of the photocatalysis lamp (Vis + UV-A) marking ZnO as the only photoactive phase in obtained coatings, but performance of the ZnO rich coatings is not as expected. Unfortunately, the amount of ZnO decreases with increasing voltage and prolonged PEO processing time, which both favour the formation of zinc-aluminate spinel with bandgap in UV-B/C region [70].

However, the photoactivity correlates with the other properties (wear and corrosion) suggesting that it is mainly the porosity and surface roughness which is the influencing factor for the properties of the coatings although the effects can be opposite depending on the property of interest. Higher photoactivity for the specimens produced at 125 mA/cm$^2$ is observed but worse corrosion and wear resistance because of the less perfect surface structure and higher internal porosity. For the specimen produced at 150 mA/cm$^2$ it is exactly the opposite, having low photoactivity but better wear and corrosion properties. This indicates that the intended repair process of the coating was at least partly succesful, but is more complex than expected. The repair mode uses current and voltage limits of the power supply [51]. Thus, with higher current density the voltage limit should be reached faster and longer time for the repair with reduced current density should be available, reaching in the best case zero current. However, this is the case for the coatings produced at 150 mA cm$^{-2}$, but not for the 125 mA cm$^{-2}$ which is performing worse compared to the 100 mA cm$^{-2}$. This may indicate that too high coating growth rate, driven by the current density, produces less dense or defect free coatings which need more repair. The 125 mA cm$^{-2}$ coatings appear to have no proper balance between coating growth and repair and thus often have the poorest properties (except photo-activity). Summarising, it is obvious that some properties (photo-activity) do benefit from rougher and more porous layers while other properties (wear and corrosion resistance) are showing negative effects on the performance and a general guideline to improve all properties is not existing.

**Conclusions**

In previous work we have demonstrated that composite $ZnO/ZnAl_2O_4$ coatings can be produced on Zn substrates by PEO processing in an alkaline phosphate-aluminate based electrolyte [51]. In the present study the focus was on the properties of those layers and the following conclusions can be drawn:

1) The introduced repair mode [51] of the PEO process to improve adhesion/cohesion of the coatings was only partly successful.

2) The coatings suffer from poor cohesion within the layer (around 1 MPa in pull-off tests). A possible explanation is the sponge-like structure with a high volume of micro- and nano-porosities in combination with poor sintering by the discharges.
3) In spite of the sponge-like structure and low internal layer cohesion, the wear and corrosion properties are reasonable, much better than the untreated substrate, thus industrial and biomedical applications are still feasible.
4) The PEO coatings is preventing a strong adhesive metal/metal fretting wear with enhanced transfer of Zn metal to the steel counter part by changing the wear mode to a mild abrasive wear affecting both coating and ball.
5) The porous structure of the coatings is detrimental for the corrosion resistance, but offers also interesting alternatives. Closer investigation reveals a kind of self-healing ability which is based on blocking of the pores by corrosion products. In industrial applications such a self-healing ability can be supported by proper selection and infiltration of corrosion inhibitors into the coating and post-sealing by paints with the advantage to improve also the mechanical stability of the composite layer system. In a similar approach drugs can be incorporated for biomedical applications and sealed with degradable applications.
6) The photoactivity is not good because of the band gap of 3.9 eV for the main $ZnAl_2O_4$ phase of the coatings which does not allow excitation in either the visible or in UV-A light range. Thus applications for water purification are less likely. Nevertheless, there is some activity present which may at least support UV sterilisation in biomedical applications.

Summarising, reasonable improvements in wear and corrosion resistance compared to the Zn substrate suggest more the classical industrial applications but also biomedical applications to control the degradation rate are an option.

**Acknowledgements**

This work was partially supported by REA in frame of Horizon2020-MSCA/RISE-2018. Nr. 823942 (FUNCOAT project: "Development and design of novel multiFUNctional PEO COATings").

**Conflict of interests**

No conflict of interests can be indicated.